\documentclass[useAMS,usenatbib,usegraphicx]{mn2e}
\usepackage[draft]{hyperref} % required for mnras.bst
\usepackage[modulo]{lineno} % the modulo causes only one in 5 to be printed
\usepackage{xfrac} %  allows for nice fractions in text 
\usepackage[svgnames]{xcolor}  %

\newcommand{\arcdeg}{\degr}

\newcommand{\Msolxyr}{\mbox{$M_{\sun}$~yr$^{-1} / 1000$~km~s$^{-1}$}}

\newcommand{\kms}{\mbox{km s$^{-1}$}}
\newcommand{\muas}{\mbox{$\mu$as}}

\newcommand{\muasd}{\mbox{$\mu$as~d$^{-1}$}}

\newcommand{\Jb}{\mbox Jy~beam$^{-1}$}
\newcommand{\muJb}{\mbox{$\mu$Jy~beam$^{-1}$}}
\newcommand{\RA}[4]{\mbox{${#1}^{\rm h} \; {#2}^{\rm m} \; {#3}\fs{#4} $}}
\newcommand{\dec}[4]{\mbox{${#1}\arcdeg \; {#2}\arcmin \; {#3}\farcs{#4} $}}
\newcommand{\AT}{\mbox{AT\,2018cow}}
\newcommand{\JREF}{\mbox{J1619+2247}}

\newcommand{\lesssim}{\mbox{\raisebox{-0.3em}{$\stackrel{\textstyle <}{\sim}$}}}
\newcommand{\gtrsim}{\mbox{\raisebox{-0.3em}{$\stackrel{\textstyle >}{\sim}$}}}
\newcommand{\tablenotemark}[1]{$^{\mathrm #1}$}
\newcommand{\tablenotetext}[2]{\noindent$^{\mathrm #1}$ #2\\}
\newcommand{\phn}{\phantom{1}}

\newcommand{\slugcomment}[1]{\date{#1}}

\newcommand{\aap}{Astron.\ Astrophys.}

\newcommand{\aj}{AJ}
\newcommand{\apj}{ApJ}
\newcommand{\apjl}{ApJL}

\newcommand{\mnras}{MNRAS}

\newcommand{\nat}{Nat}
\newcommand{\pasa}{PASA}

\newcommand{\pasp}{PASP}

\title{AT\,2018cow VLBI: No Long-Lived Relativistic Outflow}

\author[Bietenholz et al]{Michael F. Bietenholz$^{1,2}$, 
Raffaella Margutti$^{3,4}$,
Deanne Coppejans$^3$, \and
Kate D. Alexander$^{3,5}$,
Megan Argo$^{6,7}$, 
Norbert Bartel$^2$, 
Tarraneh Eftekhari$^8$,
\and Dan Milisavljevic$^{9}$, % he seems to use both Dan and Danny in print
Giacomo Terreran$^3$ and Edo Berger$^8$
\\
$^1$Hartebeesthoek Radio Observatory, PO Box 443, Krugersdorp,
1740, South Africa \\
$^2$Department of Physics and Astronomy, York University, Toronto,
M3J~1P3, Ontario, Canada \\
$^3$Center for Interdisciplinary Exploration and Research in Astrophysics (CIERA) and Department of Physics and Astronomy,\\
Northwestern University, Evanston, IL 60208, USA \\
$^4$CIFAR Azrieli Global Scholar, Gravity \& the Extreme Universe Program,
2019\\
$^5$NASA Einstein Fellow \\
$^6$Jeremiah Horrocks Institute, University of Central Lancashire, Preston, Lancashire PR1 2HE, UK \\
$^7$Jodrell Bank Centre for Astrophysics, Department of Physics and Astronomy, University of Manchester, M13 9PL, UK\\
$^8$Harvard-Smithsonian Center for Astrophysics, 60
Garden Street, Cambridge, MA 02138, USA \\
$^{9}$Department of Physics and Astronomy, Purdue University, 525 Northwestern Ave., West Lafayette, IN 47907, USA 
}

\begin{document}

%\slugcomment{Version 4.3, \today}
\slugcomment{Accepted for publication in {\em MNRAS}}

\pagerange{\pageref
{firstpage}--\pageref{lastpage}} \pubyear{2019}
\maketitle
\label{firstpage}

%\linenumbers

\begin{abstract}
  We report on VLBI observations of the fast and blue optical
  transient (FBOT), \AT\@.  At $\sim$62~Mpc, \AT\ is the first
  relatively nearby FBOT\@.  The nature of \AT\ is not clear, although
  various hypotheses from a tidal disruption event to different kinds
  of supernovae have been suggested.  It had a very fast rise time
  (3.5~d) and an almost featureless blue spectrum although high
  photospheric velocities (40,000~\kms) were suggested early on. The
  X-ray luminosity was very high,
  $\sim 1.4 \times 10^{43}$~erg~s$^{-1}$, larger than those of
  ordinary SNe, and more consistent with those of SNe associated with
  gamma-ray bursts. Variable hard X-ray emission hints at a long-lived
  ``central engine.''  It was also fairly radio luminous, with a peak
  8.4-GHz spectral luminosity of
  $\sim 4 \times 10^{28}$~erg~s$^{-1}$~Hz$^{-1}$, allowing us to make
  VLBI observations at ages between 22 and 287~d.  We do not resolve
  \AT.  Assuming a circularly symmetric source, our observations
  constrain the average apparent expansion velocity to be $<0.49\,c$
  by $t = 98$~d ($3\sigma$ limit).  We also constrain the proper
  motion of \AT\ to be $<0.51\,c$\@.  Since the radio emission
  generally traces the fastest ejecta, our observations
  make the presence of a long-lived relativistic jet
  with a lifetime of more than one month very unlikely.
\end{abstract}

\begin{keywords}
Supernovae: individual (AT 2018cow) --- radio continuum: general
\end{keywords}

\section{Introduction}
\label{sintro}

With the increasing cadence of optical surveys, an increasing number
of rapidly-evolving transients are being detected
\citep[e.g.][]{Drout+2014, Tanaka+2016, Pursiainen+2018}.  These rapid
transients form a diverse population, spanning a wide range of
luminosity, composition and environment, and both broaden and
challenge our current ideas of core-collapse stellar death.

\AT\ (also known as ATLAS18qqn, SN~2018cow) is in the star-forming dwarf spiral
galaxy
CGCG~137$-$68 (also known as CGCG 1613.8+2224 and SDSS
J161600.57+221608.2) at $z=0.04145$ \citep{Prentice+2018, Smartt+2018a},
which corresponds to a luminosity distance, $D_{\rm Lum} = 64$~Mpc and
an angular size distance $D_{\rm Ang} = 62$~Mpc\footnote{We use the
  values from the \citet{Planck+2018vi}, which are
  $H_0 = 67.4$~\kms~Mpc$^{-1}$, $\Omega_{\rm matter} =0.315$ and
  $\Omega_\Lambda = 0.685$.}.
\AT\ is one of a new class of fast and blue optical transients
\citep[FBOTs\footnote{Some authors use the term Fast-Evolving Luminous
  Transient, or FELT instead of FBOT.}; e.g.][]{Drout+2014}, and is
the first example of a FBOT seen in the local universe.

\AT\ was initially optically detected by the Asteroid
Terrestrial-impact Last Alert System (ATLAS) survey on MJD 58285.44
\citep{Smartt+2018a}.  It was not detected by the All Sky Automated
Survey for SuperNovae (ASAS-SN) on MJD 58284.13 \citep{Prentice+2018},
therefore the explosion date is tightly constrained, and we take a
rounded value of MJD 58285 (2018 June 16) as our explosion time,
$t = 0$ \citep[also adopted by ][]{Ho+2019a, Perley+2019}

\AT\ was also detected in the radio, first at mm-wavelengths
\citep{deUgartePostigo+2018, Ho+2019a}, 
then with the Arcminute Microkelvin Imager Large Array at 15~GHz
\citep{Bright+2018}, and subsequently with the Australia Telescope
Compact Array and the Jansky Very Large Array (VLA) at various
frequencies between 1.3 and 34 GHz \citep{Dobie+2018a, Dobie+2018b,
  Dobie+2018c, Margutti+2019_AT2018cow}.  \citet{Horesh+2018} reported
a 5-GHz detection with e-Merlin which provided a position accurate to
a few mas.
We detected it at 22~GHz with Very Long Baseline Interferometry (VLBI)
on 2018 Jul.\ 7 with the National Radio Astronomy Observatory (NRAO)
High Sensitivity Array \citep{AT2018cow_Atel}, refined the position to
the sub-mas level, and found a total flux density of $\sim$5~mJy at
22~GHz \citep{Margutti+2019_AT2018cow}.  It was subsequently also
detected with the European VLBI Network at 1.6~GHz \citep{An2018}.

\AT\ has extremely peculiar properties, which make the identification
of its intrinsic nature a challenge.  It had:
\begin{trivlist}
\item{(1)} A very rapid rise in the optical lightcurve,
brightening by 5 mag in a few days \citep{Smartt+2018b}, to a large
peak bolometric luminosity of $\sim 4 \times 10^{44}$~erg~s$^{-1}$,
followed by a relatively quick decay with luminosity declining
approximately as $t^{-2.5}$ \citep{Perley+2019}.  The high luminosity,
quick rise and rapid decay rule out optical emission powered by the
decay of $^{56}$Ni such as that in most SNe
\citep{Margutti+2019_AT2018cow}.
\item{(2)} Persistently blue colours, with an initially almost
  featureless spectrum, although some transient lines with a width of
  $\sim 0.3\,c$ were seen between $t = 4$ and 8~days \citep{Izzo+2018,
    Xu+2018, Perley+2019};
\item{(3)} Emission lines of H and He of intermediate width (a few
  thousand \kms) appeared after about 10 days, which were initially
  quite asymmetric and shifted towards the red, but which became
  more symmetric and moved blueward at later times
  \citep{Perley+2019}.
\item{(4)} An X-ray luminosity with a high peak of
  $\sim 3 \times 10^{43}$~erg~s$^{-1}$, which subsequently decayed
  rapidly \citep{RiveraSandovalM2018a, Margutti+2019_AT2018cow}.  The
  peak X-ray luminosity is comparable to those of supernovae (SNe)
  connected to gamma-ray bursts \citep[GRBs, see
  e.g.][]{DwarkadasG2012}, but larger than that of most ordinary SNe.
  The decay rate of the X-ray flux increased after $t \sim 20$~d.  In
  addition to the overall rise and decay, the X-ray emission showed
  variability with timescales as short as 1 day
  \citep{Kuin+2019,Margutti+2019_AT2018cow}.
\item{(5)} A relatively high radio luminosity, with a peak
  $L_{\nu = 8.5 {\rm GHz}}$ of $\sim 4 \times 10^{28}$~erg~s$^{-1}$~Hz$^{-1}$,
  % from Margutti+2019: nuLnu = 4.3e38 erg/s for D_Lum=64 Mpc
  %  22 GHz           t_Margutti = 21.6d  -> 5.85 mJy -> 2.97e28 erg/s/Hz, nuLnu=6.39e38 erg/Hz
  %  8.3 GHz (interp) t_Margutti = ~83d   -> 9.33 mJy -> 4.57e28 erg/s/Hz, nuLnu=3.80e38 erg/Hz
  %   
  higher than most Ibc SNe, but comparable to GRBs in the local
  universe.  In the radio, the rise was relatively slow, with the peak
  time at 8.5~GHz not occurring till $t = 80$~d
  \citep{Margutti+2019_AT2018cow}. The radio spectral energy
  distribution (SED) showed a spectral peak at $\sim$120~GHz at
  $t = 10$~d \citep{Ho+2019a}, which moved downwards in frequency to
  $\sim$5~GHz by $t = 132$~d \citep{Margutti+2019_AT2018cow}.
\end{trivlist}

Multi-wavelength observations have shown evidence for strong
asymmetries in the ejecta of AT2018cow
\citep{Margutti+2019_AT2018cow}.  Various different ejecta velocity
regimes have been observed in \AT.  The early broad spectral features
suggested some velocities of $\gtrsim 0.3 \, c$.  The radio spectral
energy distribution suggests velocities of $\gtrsim 0.1 \, c$.
Finally the H and He spectral features which emerged later suggest
velocities of $\sim 0.02 \, c$.

Spectropolarimetry at $5\leq t \leq8$~d showed significant time- and
frequency-dependent linear polarization, which is usually interpreted
as indicating significant departures from symmetry \citep{Smith+2018},
and suggesting the possibility of a segmented, anisotropic outflow of
some kind.

In fact, various authors have already suggested that there might be a
jet in \AT\ \citep{Margutti+2019_AT2018cow, Kuin+2019, Perley+2019,
  SokerGG2019}.  Due to the similarities with GRB-SNe, an off-axis GRB
event, with a relativistic jet not directed along the line of sight,
is a possibility.  No gamma-ray emission was seen to limits of
$3\times10^7$~erg~cm$^{-2}$ for a 10-s bin \citep[and references
therein]{Margutti+2019_AT2018cow, Kuin+2019}.
The X-ray data, however, suggest some form of energy input
\citep{Margutti+2019_AT2018cow, Fang+2019}.

As shown in \citet{Margutti+2019_AT2018cow}, GRB-like relativistic
jets with isotropic-equivalent energies $E_{\rm iso} \geq 10^{52}$~erg
and expanding in a wind-stratified medium ($\rho \propto r^{-2}$) are
excluded by the observations for all viewing angles for progenitor
mass-loss rates, $\dot M > 10^{-4}$~\Msolxyr.
Jets with lower $E_{\rm iso}$ or lower $\dot M$ are possible for a
range of viewing angles.

Alternatively, \citet{SokerGG2019} interpret \AT\ as the result of a
binary star where a neutron-star inspirals into its red giant
companion, accreting rapidly when it reaches the dense core.  Jets are
produced, which clear the polar regions of the supergiant, which then
form the observed high-velocity material.

Finally, \citet{Perley+2019}, \citet{Michalowski+2019} and
\citet{Kuin+2019}
all suggest the possibility that \AT\ might not be a core-collapse SN,
but rather a tidal disruption event (TDE), where a star is disrupted
by an intermediate-mass black hole which resides in the outskirts of
CGCG 137$-$68\@.
Some tidal disruption events (TDEs) can produce radio-bright
relativistic jets, for example Swift 164449.3+573451
\citep{Berger+2012_SwiftJ1644, Zauderer+2011}, although Swift
164449.3+573451 was much more radio luminous than \AT.  Recently,
\citet{Mattila+2018} reported on a TDE with a resolved relativistic
jet, Arp 299-B~AT1, which had a peak radio luminosity
$\nu L_\nu \sim 6 \times 10^{38}$~erg~s$^{1}$ \citep{Mattila+2018}
comparable to that of \AT\
\citep[$\sim 4 \times
10^{38}$~erg~s$^{-1}$;][]{Margutti+2019_AT2018cow}.  In VLBI
observations of Arp 299-B~AT1 carried out between 2005 and 2015,
\citet{Mattila+2018} found clear proper motions corresponding to a
projected speed of $\sim 0.25c$, but surmise speeds nearer to $c$ for
the first year.  The jet initially moved with a speed of near $c$, but
slowed down to $\sim 0.2 c$ after $\sim$2~yr \citep{Mattila+2018}.

Regardless of the nature of the outflow, a direct measurement of the
size of the emitting region and the expansion speed represents a key
constraint to the physics.  It is generally thought that the radio
emission in both SNe and GRB jets is produced mostly from the external
shock, that is where the ejecta impact the circumstellar or
interstellar mediums (CSM or ISM).  This means that the radio emission
is produced by the fastest ejecta. (In the case of GRB jets, shocks
internal to the jet are thought to be responsible for the short-lived
high energy emission, but the longer-lived ``afterglow'' emission at
lower photon energies, including radio, is thought to be largely due
to the external shock where the ejecta interact with the surrounding
material, e.g.\ \citealt{GranotvdH2014}; \citealt{GehrelsM2012}.) VLBI
radio observations have the unique capability of resolving the source,
and therefore represent the most direct way of observationally
constraining the size, and therefore the speed of the outflow.  Thus
motivated we undertook VLBI observations of \AT, and we present and
discuss our results in this paper.

\section{Observations and Data Reduction}
\label{sobs}

We obtained four VLBI observing sessions on \AT\ with the National
Radio Astronomy Observatory (NRAO) High Sensitivity Array (HSA), which
includes the Very Long Baseline Array (VLBA) as well as the 100-m
diameter Effelsberg telescope in Germany and the $\sim$105~m diameter
Robert C. Byrd telescope at Green Bank.  The observing runs occurred
between 2018 July and 2019 March, and we give the particulars in
Table~\ref{tobs}.

\begin{table*}
\begin{minipage}[t]{\textwidth}
\caption{VLBI Observations of \AT}
\label{tobs}
\begin{tabular}{l l l c c r c}
\hline
Date\tablenotemark{b}  &  Proposal & Telescopes\tablenotemark{c} &
Freq. & MJD\tablenotemark{d} & Age\tablenotemark{e} &
Total time\tablenotemark{f} \\
 & & & (GHz) &  & (d)~ & (h)\\
\hline
\hline
2018 Jul 7  & BB399  & VLBA except NL \& EB     & 22.3 & 58307.1 & 21.6 & 6.0 \\
2018 Aug 2  & BB401A & VLBA except PT, LA \& EB & 22.3 & 58333.0 & 47.6 & 6.0 \\
2018 Sep 23 & BB401B & VLBA \& EB               & 22.3 & 58383.8 & 98.4 & 6.0 \\
2019 Mar 30 & BB408  & VLBA, EB, GB, \& VLA     &\phn8.4 & 58572.3 & 287.3 & 8.0 \\
\hline
\end{tabular} 
\\
\tablenotetext{a}{NRAO observing code}
\tablenotetext{b}{The starting date of the observations}
\tablenotetext{c}{VLBA = NRAO Very Long Baseline Array, $10 \times
  25$~m diameter; GB = Robert C. Byrd telescope at Green Bank,
  $\sim$105~m diameter; VLA = the Jansky Very Large Array in
  phased-array mode, equivalent diameter 94~m; EB = the Effelsberg
  antenna, 100~m diameter}
\tablenotetext{d}{Modified Julian Date of midpoint of observation}
\tablenotetext{e}{The age of \AT\ since 2018 June 16 \citep{Smartt+2018a}}
\tablenotetext{f}{The total length of the observing run}
\end{minipage}
\end{table*}

We observed at 22.3~GHz, for the first three sessions, and then
switched to 8.4~GHz for the last one, recording both senses of
circular polarization over a bandwidth of 256~MHz.  As usual, a
hydrogen maser was used as a time and frequency standard at each
telescope, and we recorded with the RDBE/Mark5C wide-band system at a
sample-rate of 2~Gbps, and correlated the data with NRAO's VLBA DiFX
correlator \citep{Deller+2011a}.

The data reduction was carried out with NRAO's Astronomical Image
Processing System (AIPS).  The initial flux density calibration was
done through measurements of the system temperature at each telescope,
and improved through self-calibration of the phase-reference source,
which is an ICRF2 defining source, ICRF J161914.8+224747
\citep{Fey+2015}, or QSO J1619+2247 (just \JREF\ hereafter; we will
discuss \JREF, which turned out to be significantly resolved, in more
detail in section \ref{spm} below).

\section{VLBI images}  
\label{svlbiimg}

\begin{figure}
\centering
\includegraphics[width=0.48\textwidth]{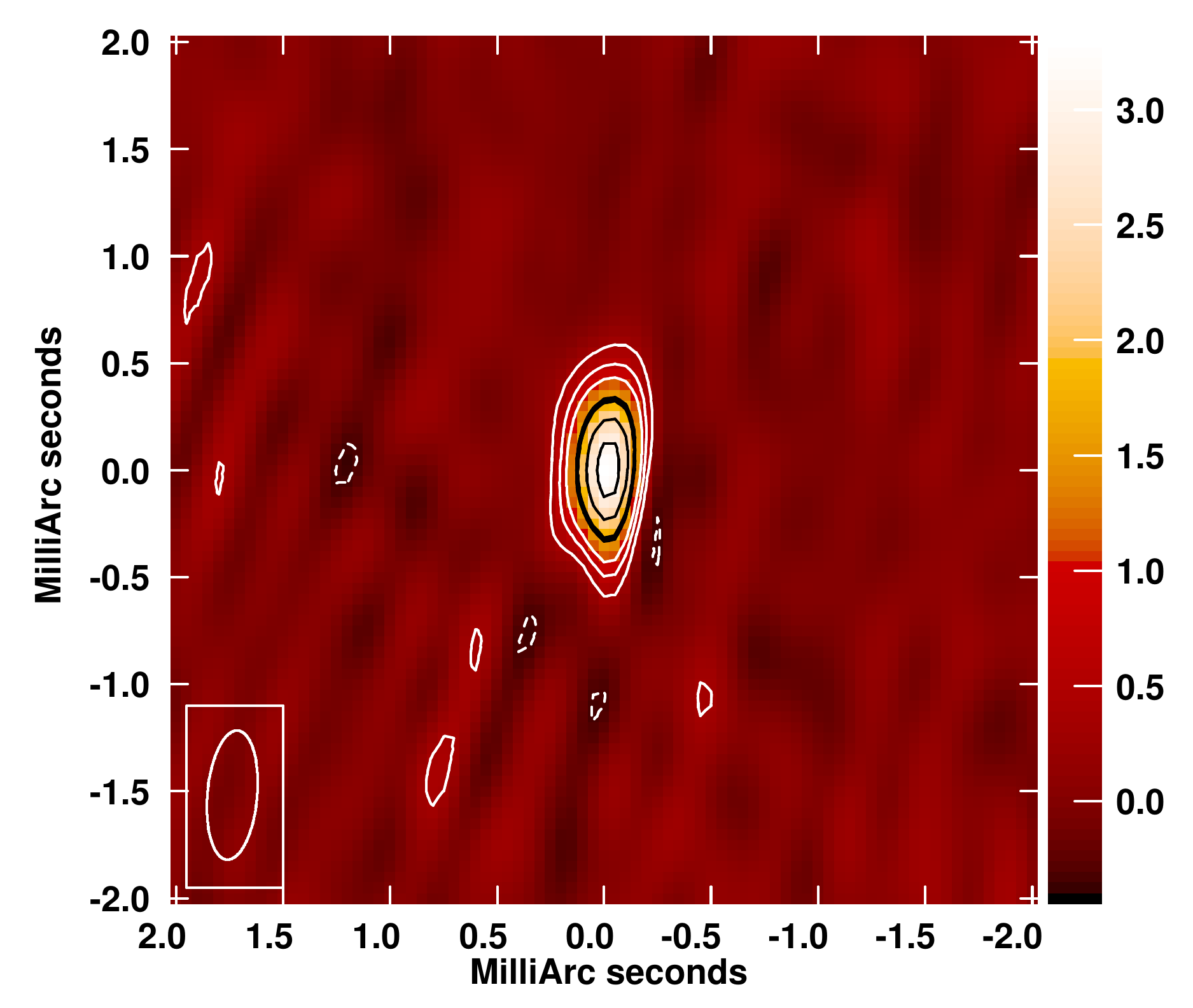}
%C  dailyalice AT2018C PS1 .ICL001.7; robust 0; BLC 279,221; trc 359,301
%C  RGBGAMMA 0
\caption{A 22.3~GHz VLBI image of \AT, observed on 2018 Sep 23\@.
  Both the contours and the colourscale show brightness.  The contours
  are at $-10$, 10, 20, 30, {\bf 50} (emphasized), 70 and 90\% of the
  peak brightness which was 3290~\muJb.  The FWHM of the restoring
  beam, which was $0.60 \times 0.23$~mas at p.a.\ $-5$\arcdeg, is
  indicated at lower left. North is up and east is to the left, and we
  take the peak-brightness point as the origin of the coordinate
  system.  The rms background brightness was 111~\muJb.}
\label{fimg}
\end{figure}

In Figure~\ref{fimg}, we show one of our VLBI images of \AT, at
22.3~GHz and observed on 2018 September 23, at $t = 98.4$~d.  Since
\AT\ is unresolved in all our observing sessions, we do not reproduce
the other images.

As we will show later, $t = 98.4$~d is also the epoch for which we
obtain the most stringent quantitative constraint on the size of the
radio source. The image appears to be largely if not completely
unresolved with the 50\% contour being very similar to that of the
restoring beam.

Other than the central peak of \AT, no emission was seen that was
brighter than $440 \, \muJb$, or $<13$\% of \AT's peak brightness
(note that only a portion of the imaged area is reproduced in
Fig.~\ref{fimg}).  We can therefore say that, between the radii of
0.6~mas (our beamwidth) and 50~mas, we can see no emission which could
be the result of a highly relativistic jet to those limits.  The range
of radii correspond to projected speeds of $2.2\,c$ to $>100\,c$.
  %away from the brightness peak, which would correspond to apparent
  %(projected) speeds of $2.2c$ to $>100$~c, must be below this limit.}
  %  Deanne's suggestion here
  % my orig:

\AT\ is also largely or completely unresolved in our images at other
epochs, and in no case is any significant emission displaced from the
central peak seen.  We chose not to reproduce the other images in this
paper since the source is unresolved.

Of our four images, \AT's flux density was highest at $t = 47.6$~d and
22.3~GHz, and this image also has the highest dynamic range.  In this
image, any emission at separations between 0.6~mas to 50~mas from the
brightness peak, which would correspond to apparent speeds of $4.5\,c$
to $>200\,c$, must be $<1.1$~m\Jb, or $<6.5$\% of \AT's peak
brightness.  On the $t = 21.6$~d, 22.3~GHz image, there is no emission
displaced from the peak of \AT\ $> 1.1$~m\Jb, and on the $t=287.3$~d
image, none $> 16 \, \mu$\Jb.
  
\section{Size, Expansion Speed and Proper Motion}
\label{ssize}

AT~2018cow is unresolved in all our VLBI observations.  In all cases a
point source is compatible with our measurements.  However, we would
like to place some upper limits on its angular size. This can be done
most accurately by fitting models directly to the visibility
(Fourier-transform plane) data, which generally permits higher
accuracies than fitting the image data \citep[see][for more detailed
discussions of this process]{SN2008D-VLBI,SN93J-2}.  To do so,
however, requires the assumption of some sort of model geometry.

In a normal SN, an approximately spherical outflow produces a forward
and reverse shock structure, with the radio emission arising in the
region in between, which is expected therefore to have an
approximately spherical-shell geometry.  In the earlier stages when
the emission is optically thick, the radio emission region would
therefore be approximately disk-like on the sky, while a more
``doughnut-like'' pattern is produced after it becomes optically thin.
Indeed, the relatively few SNe that have been resolved show structures
at least approximately like this \citep[see, e.g.][]{SNVLBI_Cagliari}.

In the case of a directed outflow like a jet, the situation is more
complicated, and a wide variety of emission geometries is possible
depending on the outflow speed, opening angle and the angle between
the jet axis and line of sight.  Although one might naively expect the
radio emission to be elongated along the projected jet axis,
\citet{GranotDR2018} calculated model radio images for various GRB
jets, and depending on the time and other parameters, a wide variety
of radio morphologies were produced. For example, the radio emission
could be elongated perpendicular to the jet direction, but displaced
from the explosion center (bow shock), or elongated along the jet
direction when both jet and counterjet are
visible. \citet{GranotDR2018} found significant proper motion of the
radio emission centroid in many cases.  Similar results are seen by
\citet{WuM2018, WuM2019} who calculated models for an off-axis jet in
the binary neutron-star merger event GW~170817, and found that even an
initially highly-directed outflow rapidly becomes extended in the
direction perpendicular to the jet axis, although the emission region
may be displaced from the explosion center along the jet axis.
We therefore expect that by the time the radio emission becomes
bright, the shock structure is already significantly sphericized, and
in projection is more likely to be more circular, or possibly
bow-shock shaped, rather than highly elongated along the jet
direction.  (We discuss the proper motion in the case of \AT\ in
\S~\ref{spm} below).

Given the range of possible geometries for \AT, and our lack of
resolution, we restrict ourselves to the one simple model which can
give some representative results for the various possible real
geometries.  We choose a circular disk model, which is bounded and
therefore provides a convenient estimate of the outer radius of the
emission region.  As mentioned above, such a disk resembles the
expected emission in the case of a young, optically-thick SN, and
should fairly representative results in other cases.  We discuss the
effect of our choice of model on our estimates of the expansion speed
below.

\begin{figure}
\centering
\includegraphics[width=0.48\textwidth]{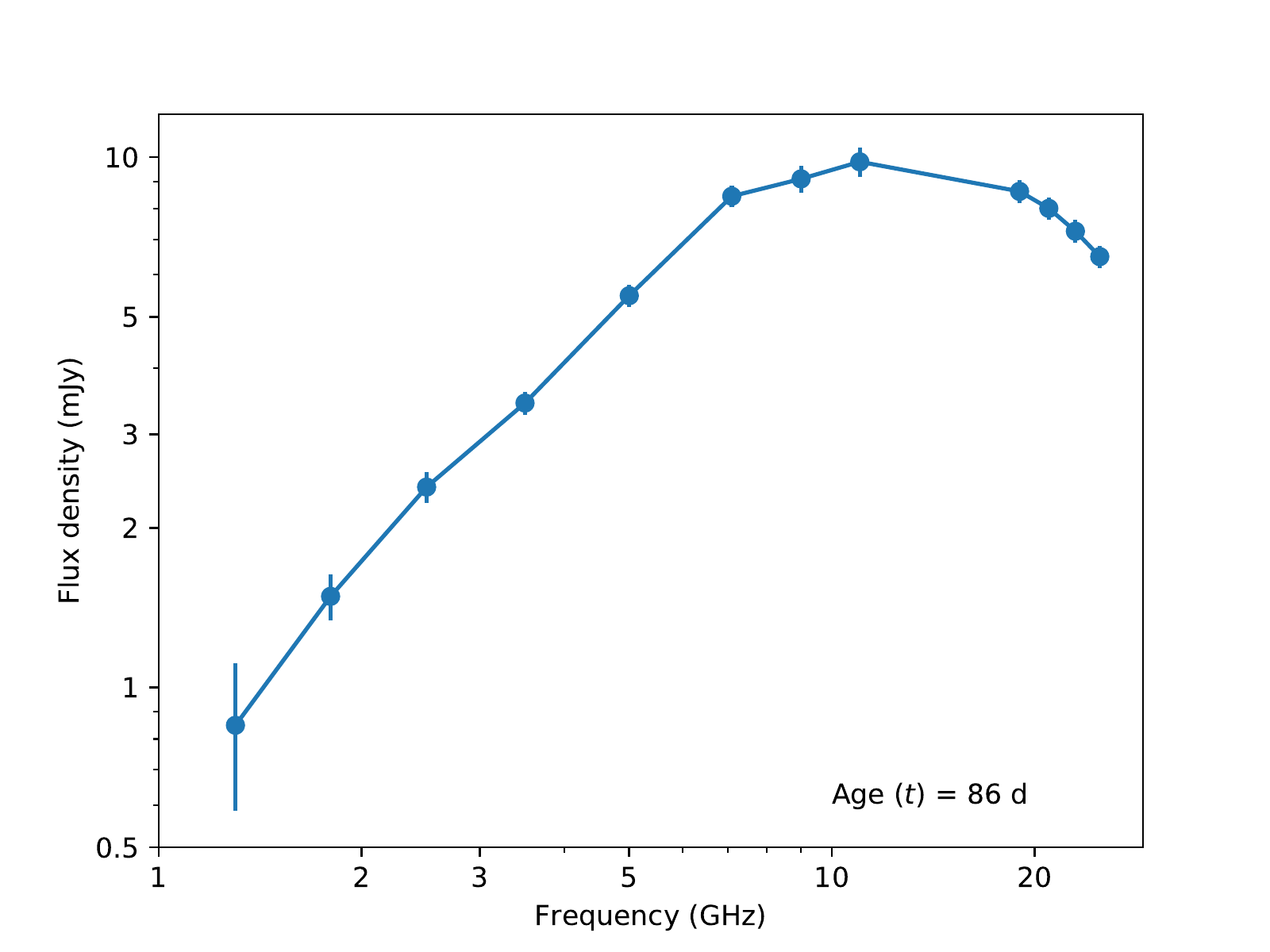}
\caption{An example radio SED (spectral energy distribution) for
  AT~2018cow at age $(t) \simeq 86$~d.  The data are taken from
  \citet{Margutti+2019_AT2018cow}, and were observed between
  $t = 83$ and 92~d. The plotted uncertainties include a 5\%
  systematic contribution from the uncertainty in the flux-density
  bootstrapping.  At this $t$, the spectral peak is near 12~GHz,
  implying that at this time the source is optically thin at
  frequencies well above 12~GHz, and optically thick at frequencies
  well below.}
\label{fSED}
\end{figure}

We show an example of \AT's SED, at $t \simeq 86$~d in
Figure~\ref{fSED}.  At this time the spectral peak was near 12~GHz,
implying that our 22~GHz observations at $t = 98.4$~d, while
nominally optically thin, were still in the transition region between
the optically thick and thin regimes.  Our observations at 8.4~GHz and
$t = 287.3$, were well in the optically-thin regime.  If the emission
region is in fact a spherical shell as expected for a normal
supernova, a spherical shell model (such as we used in e.g.,
\citealt{SN93J-3} and \citealt{Bietenholz+SN2011dh-III}) would be more
appropriate than the disk.  However, the effect using a disk model on
our upper limits on the expansion size is small: the fitted outer
radius, or limit thereon, for the shell model would be only $\sim$3\%
smaller than the value we give.

The fits were done with the AIPS task OMFIT.  When the
signal-to-noise ratio permitted, we fitted also the antenna-gain
phases, in other words simultaneously model-fitting and
self-calibrating.

We give the fitted total flux densities in
Table~\ref{tastrometry}. Given the rapid variability of \AT, these are
reasonably consistent with those seen with the VLA and other
telescopes.  A future paper, Coppejans et al., will discuss the
multi-frequency lightcurve in more detail.

As to the outer angular radius, in all cases, only upper limits could
be determined.  In Table~\ref{tastrometry} we give the $3\sigma$ upper
limits on the outer angular radius for each of our four epochs, along
with the implied limits on the expansion speed (calculated for
$D = 62$~Mpc).  The $3\sigma$ upper limit on the expansion speed for
our last epoch at $t = 287$~d was $0.74\,c$.  The most constraining
$3\sigma$ upper limit on the angular size was that from our third
epoch, $t = 98.4$~d, which was 128~\muas, corresponding to a limit on
the average expansion speed over the first 98.4~d of $<0.49\,c$.  We
note that these limits were derived based on a model with circular
symmetry in the sky plane. If the expansion were one-sided, or the
source elongated along the N-S direction where our resolution is
poorer, then expansion speeds up to factor of $\sim$2 higher than the
values given in Table~\ref{tastrometry} could be compatible with our
measurements.

\begin{table*}
\begin{minipage}[t]{\textwidth}
\caption{Model fit results: flux density, radius and position}
\label{tastrometry}
\begin{tabular}{l c r r c c c l l}
\hline
  Date & MJD\ & Age & Flux~~ 
  & Outer Angular &
  & Expansion Velocity\tablenotemark{c}
  & \multicolumn{2}{c}{Relative position\tablenotemark{d}}\\
  %   second line
  & & & Density\tablenotemark{a} & Radius\tablenotemark{b} \\
  % third line
       &     & (d)  &  $\mu$Jy~~~ &($3\sigma$ limits; \muas)   & ($3\sigma$ limits; $v/c)$
  & RA (\muas) & decl. (\muas)       \\
\hline\hline
2018 Jul 7  & 58307.1 & 21.6  & 5870 & $<111$      & $<1.84$ &$\phn\phn0\pm66$&$-25\pm66$\\
 2018 Aug 2  & 58333.0 & 47.6 & 20100 &  $<\phn87$ & $<0.65$ &$-24\pm66$&$-42\pm65$\\
 2018 Sep 23 & 58383.8 & 98.4  & 4050 & $<134$     & $<0.49$ &\phn$46\pm66$&$-15\pm66$\\ 

 2019 Mar 30 & 58572.3 & 287.3 &  69  & $<630$     & $<0.79$ &$-22\pm70$&$\phn82\pm100$\\
\hline
\hline
\end{tabular}
\\
\tablenotetext{a}{The total flux density of a uniform circular disk
  model fitted to the calibrated visibility data by least squares}
\tablenotetext{b}{The angular outer radius of the fitted circular disk
  model.  Angular sizes larger by a factor of $\sim$2 are compatible
  with our measurements if the source is elongated primarily in the
  N-S direction where our resolution is poorer.}
\tablenotetext{c}{The average expansion speed assuming two-sided
  expansion, radius / time, taking a distance of 62~Mpc}
\tablenotetext{d}{See text, \S~\ref{spm}.  The positions given
  relative to the mean centre position of \AT\ over our four epochs,
  which was RA = \RA{16}{16}{0}{22417609}, decl.\ =
  \dec{22}{16}{4}{8903214} (J2000), and was determined relative to
  that of \JREF, with a correction for the average shift of the peak
  brightness position with frequency expected due to opacity effect
  between 22 and 8.4 GHz for the last epoch \citep[``core
  shift'';][]{Plavin+2019}.  The uncertainties include the statistical
  contribution, the contribution due to the uncertainty in position of
  the reference source, as well as a contribution due to the
  phase-referencing calculated according to \citet{PradelCL2006}}
\end{minipage}
\end{table*}

\subsection{Proper Motion}
\label{spm}

We determined the proper motion of \AT\ using our phase-referenced
VLBI observations to obtain differential astrometry between \AT\ and
our phase reference source, \JREF.  All our astrometric measurements
were made without any phase-selfcalibration, and used data that was
strictly phase-referenced to \JREF.

\begin{figure}
\centering
\includegraphics[width=\linewidth]{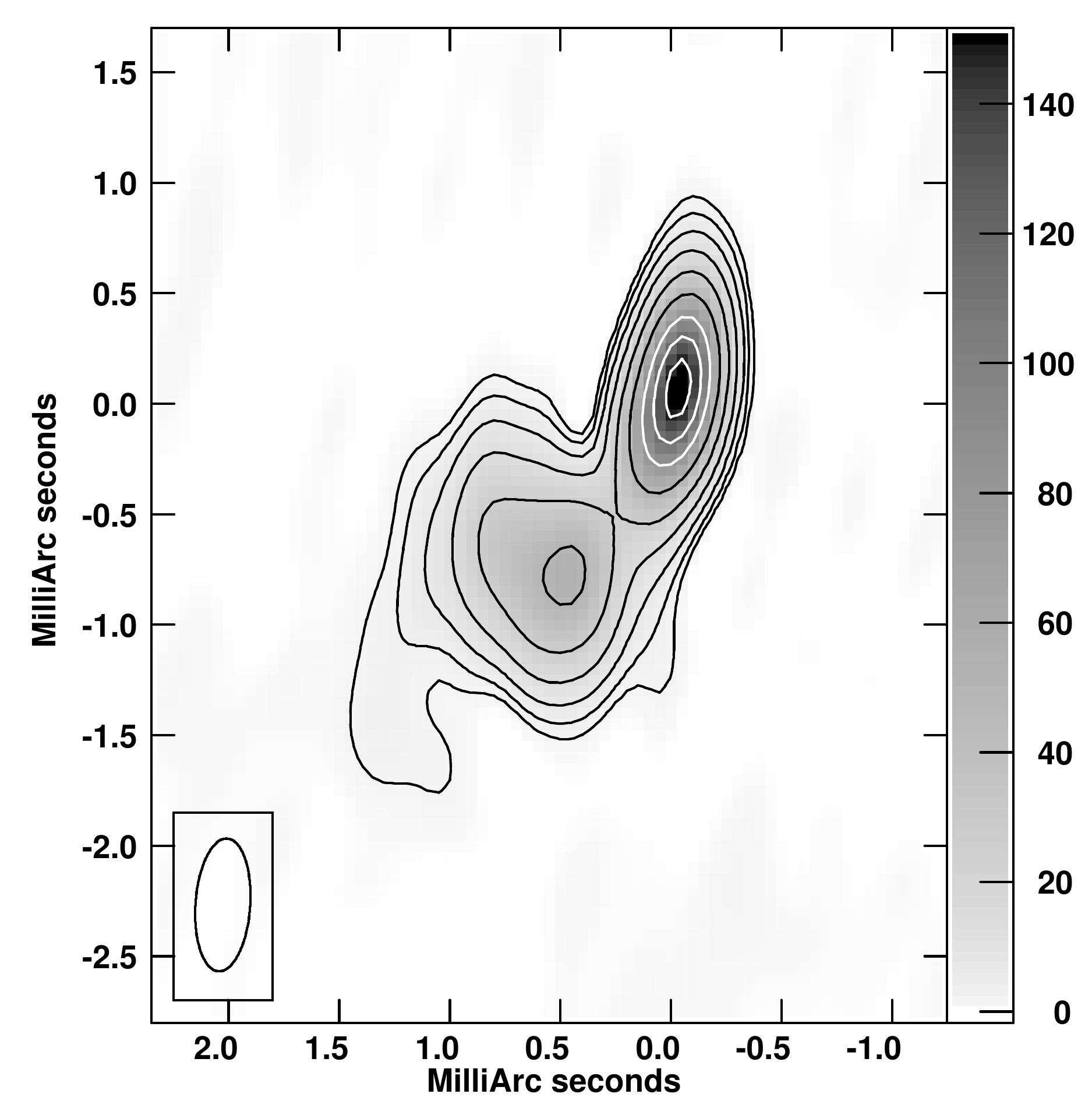}
%C dailyalice, uid=1977; J1619C C12  .ICL1LS.1
%C KNTR: blc 210,202; trc 280,290; DARKLINE = 0.48; PIXR 0 0.15;
%C LWPLA: DPARM(8)=16; LPEN=5; RGBGAMMA=0; PLVER 3; FUNCTY 'SQ';
\caption{VLBI image of our phase-calibrator source, \JREF, observed on
  2018 Sep 23 at 22.3~GHz.  Both colours and greyscale show the brightness.
  The contours are drawn at $-1$, 1, 2, 4,
  8, 16, 30, 50, 70 and 90\% of the peak brightness, which was 162
  m\Jb, with the contours at or above 50\% drawn in white.  The
  brightness scale on the right is labelled in m\Jb.  The rms
  background brightness was 0.5~m\Jb.  The FWHM of the elliptical
  Gaussian restoring beam, which was $0.60 \times 0.25$~mas at p.a.\
  $-4$\arcdeg\ is indicated at lower left.  North is up
  and east is to the left, and the origin of the coordinate system is
  the peak brightness point when convolved to the lower resolution
  available at 8.4~GHz (see text, \S~\ref{spm}).}
\label{fj1619}
\end{figure}

Our reference source, \JREF, is a ``defining'' source in the
International Celestial Reference Frame (ICRF), which is +0.75\arcdeg\
and +0.53\arcdeg\ away in RA and decl., respectively, from \AT, and
whose position is uncertain by
56~\muas\ in RA and 42~\muas\ in decl\footnote{ICRF3:
  \url{http://hpiers.obspm.fr/webiers/newwww/icrf}}.
\JREF\ is at redshift, $z = 1.99$ \citep{Sowards-Emmerd+2005}, and so
is not expected to have any discernible proper motion.  Unfortunately,
\JREF\ is not an ideal reference source as it is significantly
resolved at both of our observing frequencies.  We show the 22.3-GHz
VLBI image of \JREF\ from our 2018 September 23 epoch in
Figure~\ref{fj1619}.  The structure is likely that of one-sided jet,
with a core component and a jet or lobe component $\sim$1~mas to the
SE of core.  An elliptical Gaussian fit to the core suggests an
intrinsic (de-convolved) major axis FWHM of 350~\muas, at p.a.\
152\arcdeg.  The core component has $\sim$60\% of the total flux
density.

As a reference position, we use the position of the brightness peak of
\JREF.  Since \JREF\ is significantly resolved, the position of the
brightness peak could be resolution-dependent.  We therefore use as a
reference position that of the brightness peak on an image convolved
to our lowest resolution, that of our last epoch, observed at 8.4~GHz,
which is $0.60 \times 0.25$~mas at p.a.\ $-4$\arcdeg.  Due to
absorption effects there is still the possibility that the position of
the peak-brightness point at 22~GHz is different from that at 8.4~GHz
\citep[the ``core-shift'' phenomenon; see, e.g.][]{Kovalev+2008a}, and
therefore that our reference positions for the fourth epoch (at
8.4~GHz) is different from that used for the first three epochs (at
22.3~GHz).  \citet{Plavin+2019} found an average shift of the
peak-brightness position of 0.4~mas between 2.3 and 8.4~GHz for 40
sources, generally along the jet direction.  Assuming the magnitude of
the shift is $\propto \nu^{-1}$, and that the jet direction is p.a.\
152\arcdeg, we would expect an average shift of $\sim$80~\muas\ at
p.a.\ 152\arcdeg\ in the peak-brightness position when going from 22.3
and 8.4~GHz.  However, \citet{Plavin+2019} found that the amount of
shift varies considerably between sources, and can vary with time for
any given source, and given the complex morphology
(Fig.~\ref{fj1619}), our value for the jet direction could also be
significantly in error.  So, while a shift between 8.4 to 22~GHz of
$\sim$80~\muas\ at p.a.\ $-28$\arcdeg\ represents a ``best guess'',
the true value must be regarded as quite uncertain.

We obtained the centre position of \AT\ from similar modelfits to
those just discussed in \S~\ref{ssize} using either a circular disk,
or the projection of a spherical shell for the model.  In all cases
the position was determined without any self-calibration in phase.
The mean position over our four epochs was RA = \RA{16}{16}{0}{22418},
decl.\ = \dec{22}{16}{4}{8903} (J2000),
with an estimated uncertainty of $< 100$~\muas, which is consistent
with, but more accurate than the preliminary value we published from
only the first epoch in \citet{AT2018cow_Atel}.

We give measured offsets from the mean position in
Table~\ref{tastrometry}, with estimated standard errors.
The standard errors include three terms, all added in quadrature.
(1) the statistical uncertainties, (2) an uncertainty in the
phase-referencing, due to errors in modelling the atmospheric delay
and in the antenna positions and other components, estimated following
\citet{PradelCL2006} to be
35, 50~\muas\ in RA and decl.\ respectively for source separation
of 0.92\arcdeg, and our source declination of +22\arcdeg, and finally,
(3) the uncertainty in position of the reference source itself from
ICRF3\@. 

To obtain the proper motion of \AT, we fit a linear function to the
RA and decl.\ position offsets given in Table~\ref{tastrometry} by
weighted least-squares.  We find proper motions of
$(0.06 \pm 0.43)$~\muasd\ in RA and
$(0.44 \pm 0.23)$~\muasd\ in decl., or $0.44\pm0.33$~\muasd\ total.
Nominally, the proper motion in decl.\ is marginally significant.
However, it depends strongly on the correction for the ``core shift',
which is poorly known.  We therefore do not consider the proper motion
significant.  The formal $3\sigma$ limit on the proper motion is
1.43~\muasd, corresponding to 154,000~\kms, or $0.51\,c$.
 
\section{Discussion}
\label{sdiscuss}

\AT\ was a very unusual object, and as discussed in our introduction,
the observations in different wavelength regimes and times have
suggested an anisotropic source, with some more massive, slow ejecta
with $v \sim 0.02 \,c$, and a less massive portion with higher speeds
$v \gtrsim 0.1 \, c$.  The similarities to GRBs and the SNe associated
with them (very fast rise time and high X-ray luminosity) suggest that
there may be a relativistic component to the outflow, likely in the
form of a jet.  While jets with $E_{\rm iso} \geq 10^{52}$~erg are
disfavoured by the observations \citep{Margutti+2019_AT2018cow}, even
such energetic jets are possible at large angles to the line of sight
or for low progenitor mass-loss rates.  An off-axis GRB jet is
therefore certainly possible in \AT\@.  \citet{ChandraF2012} show that
although the majority of gamma-ray detected GRBs have
$E_{\rm iso} \geq 10^{52}$~erg, $\sim$15\% of GRBs have
$E_{\rm iso} \leq 10^{51}$~erg, so a relatively low-energy GRB is not
improbable.  The late peak in the 8.5-GHz lightcurve, at
$t \sim 100$~d, suggests an orientation not near the line of sight.

The radio SED, if due only to synchrotron self-absorption (SSA), would
suggest only non-relativistic expansion speeds of $\sim 0.1 \,c$.
There are, however, various indications of a relatively dense CSM, so
some free-free absorption (FFA) seems likely, which would make higher
expansion speeds compatible with the observed SED\@.  The spectral
index below the peak frequency is flatter ($\alpha \sim 1.3$, see
Fig.~\ref{fSED}) than expected either from SSA or FFA for a single
optical depth, suggesting a range of different optical depths is
present, which is consistent with the inferred non-spherical geometry.

The radio emission traces the fastest outflow, as it is generated in
the shocks formed where the outflow hits the CSM of ISM\@.  Our VLBI
observations placed a $3\sigma$ limit on the apparent two-sided
expansion velocity during the first 47~d of $0.65\,c$.  Our later
observations similarly rule out average expansion velocities of
$>0.49 \, c$ at $t = 98$~d (with the same caveats given for the
measurement at $t = 47$~d above).

Our upper limits on the angular expansion were based on a circular
model.  If the source were elongated along an approximately N-S
direction, or were undergoing one-sided expansion, apparent expansion
speeds of $\sim c$ would be compatible with our measurements.

Our upper limit on the proper motion, by contrast, is largely
independent on the choice of a circular model.  Our measurements put a
$3\sigma$ upper limit corresponding to $0.51 \, c$ on the proper
motion of the centroid of the radio emission over the first $\sim$9
months.  The simulations of off-axis GRB jets of
(\citealt{GranotDR2018}, see also \citealt{GillG2018}) show that in
most cases, the centroid of radio emission shows substantial proper
motion, often with superluminal apparent velocities.  Indeed, for bulk
motion with $v \sim 0.5\,c$, the majority of jet orientations would
produce apparent motions $> c$.  For GW170817, a double neutron star
merger with an off-axis GRB-like jet, \citet{Mooley+2018b} measured a
proper motion using VLBI, which corresponded to an apparent speed of
$\sim 4\,c$ over the first 230~d after the event using VLBI
observations.

Our upper limits for both expansion speed and proper motion are on the
{\em apparent}, not the physical, speeds.  In the case of a
relativistic jet, unless it was near the plane of the sky, the
simulations just mentioned show that it would likely exhibit
superluminal apparent speeds, in which case our measured limits would
{\em over}\/estimate the physical speeds.

We therefore think that in light of our measurements, it is unlikely
that there is any sustained jet with bulk motion $\gtrsim 0.5\,c$,
although we cannot conclusively rule it out.  A jet such as those seen
in GRBs, which typically only decelerate to non-relativistic speed
after times ($t_{\rm NR}$) of $\sim$~1 yr, is therefore unlikely.
Jets with a lifetime of 1 month or less were outside the time range of
our observations and are therefore still compatible with our
measurements.

As mentioned above, some authors have suggested that \AT\ is a tidal
disruption event.  \citet{Mattila+2018} saw a resolved jet in a
different TDE, Arp-B~AT1, for which they inferred proper motions and
expansion of the jet at projected speeds of $\sim c$ for the first
year.  Our observations of \AT\ clearly rule out such a long-lived and
fast jet.

One possibility for \AT\ is a choked jet formed in the stellar
collapse, where a relativistic jet is formed in the interior of the
collapsing star, but is choked before it emerges from the star's
surface.  Such a scenario has been invoked to explain the observations
of numerous powerful core-collapse SNe \citep{Piran+2019}, in
particular those of SN~2009bb \citep{Soderberg+2010_2009bb}, SN~2012ap
\citep{Margutti+2014b},
SN~2017iuk \citep{Izzo+2019} and SN~2018gep \citep{Ho+2019b}.  In this
scenario, a relativistic jet is formed inside the collapsing star, and
expands outwards through the (non-relativistic) SN ejecta.  The bulk
of the kinetic energy is in the SN ejecta but a significant fraction
is in the jet.  The jet is choked inside the star and transfers most
of its energy to a ``cocoon,'' which can emerge from the surface of
the star.  This cocoon has a small fraction of the ejected mass, and
typical velocities of order $0.1\,c$ \citep{Piran+2019}.  The cocoon
spreads laterally after it emerges, and eventually becomes relatively
spherical and blends with the remaining ejecta.  The cocoon is
expected to produce highly transient blue or UV continuum cooling
emission and broad absorption features which last typically a few
days.  Such a picture is broadly consistent with \AT, where indeed the
emission was very blue particularly early on, where the cocoon
emission may have contributed to the very rapid rise, where transient
high-velocity absorption features were seen \citep{Izzo+2018, Xu+2018,
  Perley+2019}, and where various lines of evidence suggest
significant asphericity \citep[see e.g.][]{Margutti+2019_AT2018cow,
  Smith+2018}.  Indeed, this scenario is similar to the one suggested
by \citet{Margutti+2019_AT2018cow}.  Although the shock fronts
associated with the cocoon would likely produce an initially
aspherical radio emission region, the velocities expected of the
cocoon ($0.1\,c$) are less than our observational limits on the
expansion velocity ($0.49\,c$ at $t \sim 100$~d).  The radio emission
from such a cocoon would be resolvable, but only in a relatively
nearby SN such as SN~1993J where the morphology of the forward shock
was discernible in VLBI images as early as $t \simeq 175$~d \citep[see
e.g.][]{SN93J-3}.

\section{Summary and conclusions}

\begin{trivlist}
  
\item{1.} We have made four epochs of VLBI observations of the
  unusual fast blue transient source \AT.
\item{2.} The source was unresolved in all of our observations.  We
  place upper limits on the angular size of $\lesssim 100$~\muas,
  which correspond to limits on the average apparent expansion
  velocities of $<1.84 \, c $ and $< 0.49 \, c$ at $t= 22$ and 98~d,
  respectively, assuming a circularly symmetric source.
\item{3.} We also measured the proper motion of \AT, and found that it
  a $3\sigma$ upper limit of $0.51 \, c$ between $t = 22$ and 287~d.
\item{4.} Our upper limits on the expansion velocity and the proper
  motion make a long-lived relativistic jet, such as those seen in
  GRBs, quite unlikely.
\end{trivlist}

\section*{Acknowledgements }

This research was supported by both the National Sciences and
Engineering Research Council of Canada and the National Research
Foundation of South Africa. The Margutti group at Northwestern
acknowledges  support by the National Science Foundation under Award
No.\ AST-1909796 and by NASA through contract
80NSSC19K0384\@. R. Margutti is a CIFAR Azrieli Global Scholar in the
Gravity \& the Extreme Universe Program, 2019.  K. D. Alexander
acknowledges support provided by NASA through the NASA Hubble
Fellowship grant HST-HF2-51403.001 awarded by the Space Telescope
Science Institute, which is operated by the Association of
Universities for Research in Astronomy, Inc., for NASA, under contract
NAS5-26555\@.
The National Radio Astronomy Observatory is a facility of the National
Science Foundation operated under cooperative agreement by Associated
Universities, Inc.  This work is partly based on observations with the
100-m telescope of the MPIfR (Max-Planck-Institut f\"ur
Radioastronomie) at Effelsberg, Germany.  This work made use of the
Swinburne University of Technology software correlator, DiFX,
developed as part of the Australian Major National Research Facilities
Programme and operated under licence.

\bibliographystyle{mnras} 
\bibliography{mybib1,AT2018cow.bib}

\label{lastpage}

\clearpage

\end{document}